\documentclass[10pt, draft]{article}

\usepackage{amssymb}
\usepackage{amsmath}
\usepackage{graphicx}
\title{Optimal Execution Trajectories.\\
Linear Market Impact with Exponential Decay.}
\author{ Igor Skachkov }

\begin{document}
\maketitle
\begin{abstract}
 Optimal execution of portfolio transactions is the essential part of algorithmic trading. In this paper we present in simple analytical form the optimal trajectory for risk-averse trader with the assumption of exponential market recovery and short-time investment horizon.
\end{abstract}

\section{Introduction}

At the end of the last century R.Almgren and N.Chriss \cite{AC2001, AC1999} and R. Grinold and R. Kahn \cite{GK2000} pioneered application of calculus of variation to the problem of portfolio liquidation. In the meantime most modern trading engines  use some modifications of that approach.

Following Almgren \& Chriss approach we determine the optimal trajectories by the mean-variance criterion and Constant Absolute Risk Aversion. Under that utility function  one dollar of gain can be trade-off for the same amount of risk measure regardless of the current portfolio value. In "classic" approach this measure is square dollars of variance. With linear dependance of market impact to trading rate the shape of trajectories doesn't depend on the size of order. One of the problems with those results is instantaneous recovery of price after temporary market impact. Market doesn't have memory and affect of the current execution doesn't depends on trading history. It is counterintuitive and is not consistent with the methodology of temporal market impact coefficients calibration by Almgren himself \cite{RA2005}. Obizaeva and Wang \cite{OW2005} obtained the analytic solution for risk-neutral agents when the influence of previous trades decays exponentially.
\section{Market Model}
\subsection {price dynamics}
We consider that proprietary traders converted their signals to the "alphas"
\[
\frac{dS}{S}=\alpha(t)dt+\sigma dW
\]
$W_t$ is a standard Wiener process; $\alpha_t$ and $\sigma_t$ are instantaneous drift and volatility.
The major simplification for short time horizon investing (characteristic time $t^*$ is up to few days) is a small change of all parameters including price
\[
(S(t)-S_0)/S_0\ll1
\]
It means that we don't distinguish between arithmetic and geometric Brownian motions and that the current price is approximately equals to it's expectation
\[
S \simeq E(S),  ~\sigma*t^*\ll 1
\]
The drift is either constant and local
\[
\alpha(t)=
\begin{cases}
\alpha_0, & 0<t<t_1\\
0, &\text{otherwise}
\end{cases}
\]
or decaying
\[
\alpha(t)=\alpha_0 exp(-\gamma t)
\]
Also for any successful trading strategy return should be at least of the same order as volatility
\[
\alpha t^* \sim \sigma \sqrt{t^*}
\]
In short time horizon we have for the price dynamics:
\begin{equation}
S-S_0=S_0 \int^t_0{\alpha}d\tau+S_0 \sigma W_t
\end{equation}
Additionally prices are affected by our trade. Traditionally impact is divided by two parts: permanent and temporary. Permanent impact is a linear function of the amount traded only, because otherwise we could move the value of particular asset cyclically buying and selling the same number of shares but with different speed. We don't include this permanent impact in our analysis for two reasons: first - doesn't depend on trading trajectories and second - typically the sizes of proprietary traders transactions don't exceed few percent of average daily volume ($A \! D \! V$) and correspondingly are negligible fractions of capitalization.  \begin{equation}
S-S_0=S_0 \int^t_0{\alpha(\tau)}d\tau+h[x]+S_0 \sigma W_t
\end{equation}
where $h[x]$ is a temporary market impact functional that in general depends on history of execution.

We will analyze in details market impact models in the next section.

We assume that the portfolio $\Pi$ is self-financing: the total return is equal to sum of cash received/paid from the trade and a change of portfolio value. Also it can be presented as return on holding portfolio of stocks. Stock's price dynamics is the only source of PnL, because interest rate on cash can be safely equal to zero for short time horizon.
\[
 d\Pi=\underbrace{d{xS}}_{stocks}-\underbrace{Sdx}_{cash}=
 \underbrace{xdS}_{return}
\]
where $S[x]$ is a functional price that depends on trajectory of trading.
Correspondingly the wealth dynamics is:
\begin{equation}
 \Delta \Pi=(S_T-S_0)x_T-\int^T_0{(S-S_0)\dot{x}}dt+\sigma S x dW_t
\end{equation}
Mean-variance utility
\[
\Phi=\int^t_0{(E(R)-\lambda Var(R)})dt
\]
is given by
\begin{equation}
\label{utility}
\Phi [x]=\int^T_0{(\alpha x+h[x]\dot{x}-\lambda (S x \sigma)^2 })dt
\end{equation}
for the single stock and
\begin{equation}
\label{vector utility}
\Phi=\int^T_0{(\alpha \cdot x+h[x] \cdot \dot{x}-\lambda (Sx)^t \Omega (Sx)})dt
\end{equation}
for the portfolio of stocks.

$R$ is an absolute asset price return in \$

$Var$ is a variance

$\Omega$ is a covariance matrix

$Sx=(S_1x_1 S_2x_2 ... S_nx_n)^t$ is a vector of dollar positions in stocks

$\lambda$ is a risk-averse parameter.

$T$ is a time horizon.
\subsection {market impact models}
Execution price is path-dependent in general through the affect by previous trading history. The difference between arrival price and execution price a.k.a implementation shortfall is supposed to be of opposite sign to direction of trade: we pay more to increase our position and get less to unwind it.
\begin{equation}
\label{temp market impact}
h[x]=-\int^t_0{f(\dot{x}(\tau)) K(t,\tau)}dt
\end{equation}
Market impact kernel $K(\tau,t)$ is assumed homogenous in volume time $K(\tau,t)=K(t-\tau)$. Initial time is set to $0$. Market impact is assumed to be linear functional of trading rate
\begin{equation}
\begin{split}
&h(t,\dot{x})=-\int^t_0 ({\tilde{\eta} K(t-\tau)\dot{x}(\tau))d\tau}\\
&\tilde{\eta}=\eta \frac{\sigma }{A \! D \! V S_0}
\end{split}
\end{equation}

\subsubsection {Almgren\& Chriss memoryless temporary impact}
The simplest form of the convolution integral kernel was proposed by Grinold and Kahn \cite{GK2000} and analyzed in details by Almgren and Chriss \cite{AC2001, AC2003}
\[
K(t-\tau)=\delta(0), \qquad Dirac's \ delta \ function
\]
Utility function (\ref{utility}) is given by
\begin{equation}
\label{AC utility}
\begin{split}
&\Phi[x]=-\int^T_0{\Psi(x,\dot{x},t)\,dt}\rightarrow max \\
&\Psi(x, \dot{x}, t)=-\alpha x+\eta \dot{x}^2+\lambda (xS\sigma)^2
\end{split}
\end{equation}
Subject to initial and terminal conditions:
\begin{equation}
\label{initial_conditions}
\begin{split}
x(0)=X_0 ,~x(T)=X_T
\end{split}
\end{equation}
Maximization of (\ref{AC utility}) is a standard calculus of variation problem. Applying The Euler equation
\begin{equation}
\label{euler}
\frac{\partial \Psi}{dx}- \frac{d}{dt}\frac{\partial \Psi}{\partial \dot{x}}=0
\end{equation}
we obtain
\begin{equation}
\label{AC euler}
\ddot{x}-k^2x=-\tilde{\alpha}
\end{equation}
Our solution is a simple generalization of Almgren \& Chriss \cite{AC2003} to the drift with exponential decay
\begin{equation}
\label{AC_solution}
\begin{split}
x\left(t\right)&=\left(X_0+a(0)\right)\frac{\sinh\left(k(T-t)\right)}{\sinh\left(kT\right)}+
\left(X_T+a(T)\right)\frac{\sinh\left(kt\right)}{\sinh\left(kT\right)}-a(t) \\
\dot{x} \left( t \right) &=\frac{dx}{dt} = -\left(X_0+a(0)\right)
\frac{k\cosh\left(k(T-t\right)}{\sinh\left(kT\right)}+
\left(X_T+a(T)\right)\frac{k\cosh\left(kt\right)}{\sinh\left(kT\right)} -\dot a (t)
\end{split}
\end{equation}
where

$k^2=\lambda/\tilde{\eta}$

$\tilde{\alpha}=\alpha/(2\tilde{\eta})$

$a(t)$ is a particular solution of equation (\ref{AC euler})

$a(t)=-\frac{\tilde{\alpha}_0}{\gamma^2-k^2}\exp(-\gamma t)$

\subsubsection {Exponential kernel}
The solution (\ref{AC_solution}) is a decent approximation for optimal trajectories on practice, but it cannot describe the market impact of a single discrete trade (it is a delta-function). The instantaneous recovery assumption is unrealistic and inconsistent with calibration procedures. To resolve those problems we have to replace delta function with some smooth kernel. The function under integral in (\ref{AC utility}) is given by:
\begin{equation}
\label{kernel utility}
\Psi(x, \dot{x}, t)=-\alpha x+\eta \dot{x}(t)\int^t_0{ \dot{x}(\tau)K(t-\tau)}d\tau+\lambda (xS\sigma)^2
\end{equation}
Calculating variations
\footnote{
\[
\begin{split}
&\delta_{\dot{x}}\Phi=\int^T_0{[\delta (\dot{x}(t)) \int^t_0{\dot{x}(\tau)K(t-\tau)d\tau}+\dot{x}(t)
\int^t_0{\delta( \dot{x}(\tau))K(t-\tau)d\tau}]dt}\\
& \text {We change the order of integration for the second integral and get}\\
&\delta_{\dot{x}}\Phi=\int^T_0{ \int^T_0{\dot{x}(\tau)K(|t-\tau|)}d\tau \delta (\dot{x}(t))dt}\\
\end{split}
\]
}
we obtain the following equation
\begin{equation}
\label{kernel euler}
2 \tilde{\lambda} x-\alpha= \dot{F} = \frac{d}{dt}\int^T_0{\dot{x}(\tau) K(|t-\tau|)d\tau}
\end{equation}
Assuming that the kernel is twice differentiable gives
\[
2\tilde{\lambda} \ddot{x}-\ddot{\alpha}= \frac{d}{dt}(\int^T_0{\dot{x}(\tau)\ddot K(|t-\tau|)d\tau}+2 \dot{x} \dot{K}_0(0))
\]
Now we plug the exponential kernel
\[
K(t-\tau)=\beta \cdot exp(-\beta (t-\tau))
\]
into generic equations and get the system:
\begin{equation}
\label{expKernel equations}
\begin{split}
& \ddot{x}-\frac{\lambda \beta^2}{\lambda + \beta^2}x = \ddot{x}-k^2 x = \frac{\beta^2 \alpha+\ddot{\alpha}}{2(\lambda + \beta^2)}\\
& \ddot{F}-\beta^2F=-2\beta^2\dot{x}
\end{split}
\end{equation}
From equations (\ref{expKernel equations}) follows that additionally to the smooth exponential terms the solution for trajectories contains jumps at the ends of time interval. The general solution for function $F(t)$ has the terms $exp(\pm\beta t)$. To get those terms we have to plug $\delta$-function for trading rate $\dot{x}$ into (\ref{kernel euler}). Therefore the general solution for the optimal trajectories is given by
\begin{equation}
\label{general solution exp kernel}
x=C_1e^{kt}+C_2e^{-kt}+D_1\mathcal{H}(t)+D_2\mathcal{H}(t-T)
\end{equation}
where $\mathcal{H}(t)$ is a Heavyside's function.
We need four equations to find arbitrary constants. Two equations are initial and terminal conditions and two additional equations follow from the requirement that $x(t)$ doesn't have $exp(\pm \beta t)$ terms. After some simple but tedious algebra we get a solution in a familiar form:
\begin{equation}
\label{nice_exp kernel solution}
\begin{split}
&x\left(t\right)=\left(X_0+a\right)B\frac{\sinh\left((k(T-t)+A\right)}{\sinh\left(kT+2A\right)}+
\left(X_T+a\right)B\frac{\sinh\left(kt\right)}{\sinh\left(kT+2A\right)}-a \\
&where\\
&A= \ln{\sqrt{\frac{\beta+k}{\beta-k}}}, \quad B=\frac{k}{\sqrt{\lambda}}
\end{split}
\end{equation}
For risk-neutral traders $(\lambda\rightarrow 0)$ and $\alpha=0$ the optimal schedule under exponential impact relaxation is a combination of two jumps and straight line between them.
\begin{equation}
\label{exp kernel lambda 0}
\begin{split}
&\lim_{\lambda \to 0} x=(X_0-\Delta X_0)\frac{T-t}{T}+(X_T+\Delta X_T)\frac{t}{T}\\
&\Delta X_0=\Delta X_T=\frac{X_0-X_T}{\beta T+2}
\end{split}
\end{equation}
Optimal trading strategy with the exponential kernel was the subject of A.Obizhaeva and J.Wang, \cite{OW2005}. They were probably were the first who pointed out to discontinuity of optimal paths at the ends of time interval and derived equation (\ref{exp kernel lambda 0}) for risk-neutral traders. Both discrete and continuous regimes of trading were considered using dynamic-programming approach.  They also obtained a solution in quadratures for more general case, but it is much more  complex then our analytical formula.


\begin{thebibliography}{9}
\bibitem{AC2001}
Almgren R. and N. Chriss (2001) Optimal Execution of
Portfolio Transactions, {\it Journal of Risk}, {\bf 3(2)}, 5--40.

\bibitem{AC2003}
Almgren, Robert, 2003, Optimal execution with nonlinear impact functions and trading enhanced risk, {\it Applied Mathematical Finance} {\bf 10}, 1–18.

\bibitem{AC1999}
Almgren, Robert, and Neil Chriss, 1999, Value under liquidation, {\it Risk}, {\bf 12} (12).

\bibitem{RA2005}
Almgren, Robert, Chee Thum, Emmanuel Hauptmann, and Hong Li, 2005, "Direct estimation of equity market impact", Risk 18(7), 58–62.

\bibitem{GK2000}
Grinold Richard, and Kahn Ronald, 2000, Active Portfolio Management,  McGraw-Hill,  New York.

\bibitem{OW2005}
Anna Obizhaeva and Jiang Wang, 2005
Optimal trading strategy and supply/demand dynamics.
{\it MIT working paper}
\end{thebibliography}
\end{document}